\documentclass[a4,12pt]{article} 

\usepackage{amsmath,amssymb}

\usepackage{graphicx}
\usepackage{bm}
\usepackage{hyperref}
\hypersetup{colorlinks,%
                 citecolor=blue,%
                 filecolor=black,%
                 linkcolor=black,%
                 urlcolor=blue}
\usepackage[tight]{subfigure}
\usepackage{float}
\usepackage{mathtools}

\usepackage[normalem]{ulem}
\usepackage[dvipsnames]{xcolor}

\newcommand{\beginsupplement}{%
        \setcounter{equation}{0}
        \renewcommand{\theequation}{S\arabic{equation}}%
        \setcounter{figure}{0}
        \renewcommand{\thefigure}{S\arabic{figure}}%
     }
\graphicspath{{Osc_images/}}
\begin{document}

\title{Tunability of the Dual Feedback Genetic Oscillator Modeled by the Asymmetry in Transcription and Translation}
\author{Yash Joshi$^{1}$, Yash K. Jawale$^{1}$ and Chaitanya A. Athale$^{1,\dagger}$, \\1: Div. of Biology, IISER Pune, Dr. Homi Bhabha Road,\\ Pashan, Pune 411008, India.\\
$\dagger$ All correspondence to be addressed to: \href{mailto:cathale@iiserpune.ac.in}{cathale@iiserpune.ac.in}}
\date{}

\maketitle

\section*{Abstract}
Oscillatory gene circuits are ubiquitous to biology and are involved in fundamental processes of cell cycle, circadian rhythms and developmental systems. The synthesis of small, non-natural oscillatory genetic circuits have been increasingly used to test fundamental principles of genetic network dynamics. 
A recently developed fast, tunable genetic oscillator by Stricker et al. \cite{Stricker2008} has demonstrated robustness and tunability of oscillatory behavior by combining positive and negative feedback loops. This oscillator combining {\it lacI} (negative) and {\it araC} (positive) feedback loops, was however modeled using multiple layers of differential equations to capture the molecular complexity of regulation, in order to explain the experimentally measured oscillations. 
We have developed a reduced model based on delay differential equations (DDEs) of this dual feedback loop oscillator, that reproduces the tunability of oscillator period and amplitude based on the concentration of the two inducers isopropyl $\beta$-D-1-thiogalactopyranoside (IPTG) and arabinose. Previous work had predicted a need for an asymmetry in  copy numbers of activator (araC) and repressor (lacI) genes encoded on plasmids. 
We use our reduced model to redesign the network by comparing the effect of asymmetry in gene expression at the level of (a) DNA copy numbers and the rates of (b) mRNA translation and (c) degradation. We find the minimal period of the oscillator is sensitive to DNA copy number asymmetry, but 
translation rate asymmetry has an identical effect as plasmid copy numbers, while modulating the asymmetry in mRNA degradation can improve the tunability of period of the oscillator, together with increased robustness to replication `noise' and influence of the host cell cycle. Thus, our model predicts experimentally testable principles to redesign a potentially more robust oscillatory genetic network.


\section{Introduction}
The ubiquity of oscillatory genetic networks suggests a central role in biology, resulting in extensive experimentally and theoretical studies as seen in case of the cell cycle clocks \cite{Newport:1984aa,Murray:1989aa,Novak:1993aa,Novak:1997aa,Ferrell:2011aa}, circadian rhythms \cite{Ishiura:1998aa,Stelling:2004aa} and developmental clocks in embryogenesis \cite{Horikawa:2006aa,Lewis:2007aa,Uriu:2009aa}. These oscillatory networks appear to have been selected for tunability and robustness as a part of the general `homeostatic' mechanisms of such physiological processes critical to living systems. However, understanding the design principles of such networks raises challenges  with due to their complexity. Increasingly, synthetic biology of small genetic networks has become an important alternative approach to gaining a fundamental understanding of fundamental principles of gene regulation driving such oscillators using small and relatively tractable genetic networks 
such as single gene negative-feedback systems \cite{Becskei:2000aa}, the three component `repressilator' \cite{Elowitz2000} and cell-free two- and three-stage gene cascades \cite{Noireaux:2003aa}. 
Based on theoretical studies of naturally occurring genetic oscillators, they have been broadly classified into either negative feedback loops or coupled positive and negative loops \cite{Ferrell:2011aa}. Indeed comparative modeling has demonstrated that while a minimal negative feedback loop network can produce oscillations, robustness and tunability are improved by the addition of other loops, which could explain the evolutionary selection of such complex networks \cite{Maeda2012}.

\begin{figure}[ht!]
\begin{center}
\includegraphics[width=0.4\textwidth]{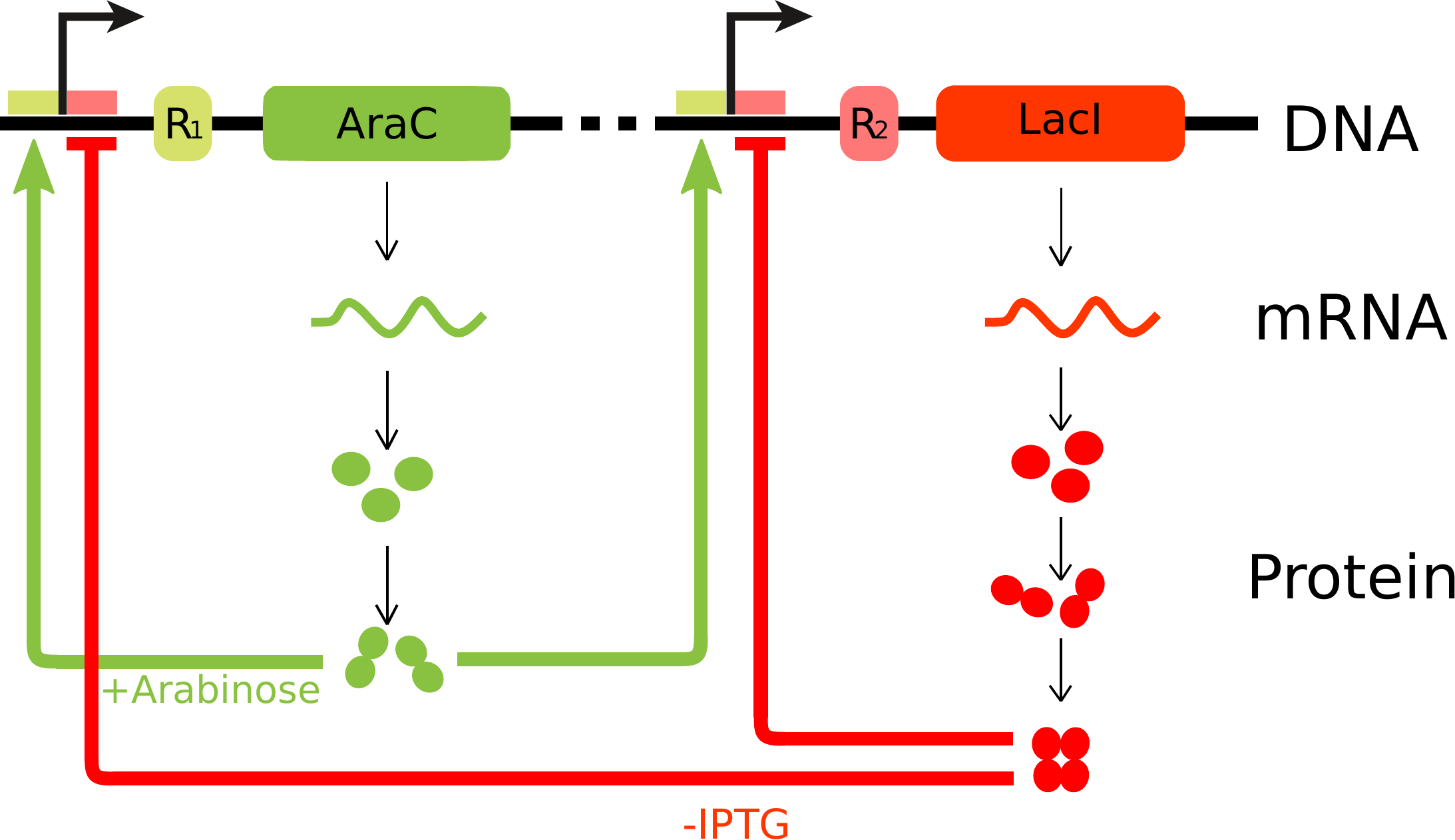}
\caption{{\bf Model of the LacI-AraC tunable genetic oscillator.} The schematic represents a kinetic model of transcriptional regulation of gene expression of \textit{araC} (green) and \textit{lacI} (red) genes by a dual-feedback loop of activation (arrowhead) by dimers of AraC proteins (green circles) and repression (bar end) by tetramers of LacI proteins (red circles). The model explicitly includes mRNA transcription, protein translation and folding and modulation of gene expression by arabinose and IPTG.}
\label{fig:schematic}
\end{center}
\end{figure}

A canonical example of a synthetic dual feedback loop - positive and negative- oscillator that has rapidly become a standard mode is the \textit{araC} and \textit{lacI} genetic oscillator with expression determined by a dual input $p_{lac/ara-1}$ promoter \cite{Stricker2008}. The genes are regulated by their own protein products (feedback) - activation by AraC protein in the presence of arabinose (positive loop) and repression by LacI protein in the absence of IPTG (negative loop). A model with twenty-seven coupled ordinary differential equations (ODEs) was developed to match the experimental findings \cite{Stricker2008}, since a simple minimal model of dual feedback loops \cite{Hasty2002}, failed to reproduce the oscillatory response of the system to parameter changes. The many intermediate reactions in the model such as the relatively slower rate of mRNA production, protein folding, protein multimerization and promoter binding play an important role in the experimental validation of this model. While explicit models of detailed molecular mechanisms are physically more realistic than minimal models, they also lead to an `explosion' of the number of parameters and variables. One solution to develop a simple model that captures the essential nature of the process, while reproducing measurable dynamics of the system is through the use of delay differential equations \cite{Yildirim:2003aa}. However, it is not clear whether a delay differential equation model could be used to describe the dual-input oscillator.

A search of general design principles of genetic oscillators points to delays and noise as important features, alongside with topology as playing an important role  \cite{Purcell2010}. The explicit use of delay differential equations for modeling genetic networks is seen in oscillator models of the cell cycle \cite{Ferrell:2011aa} and somitogenesis clock \cite{Lewis:2007aa} and lac operon dynamics \cite{Yildirim:2003aa}. Indeed a cell-free extract based study of the lac-ara dual feedback oscillator have demonstrated that protein translation can serve as a bottleneck in the dynamics of the oscillator \cite{Noireaux:2003aa}. Hence, such a separation of time-scales seen in experiment and the utility of delays in oscillatory network models, suggests such an approach could help reduce the model complexity of the lac-ara oscillator.

Here, we describe a novel model of the lactose-arabinose dual feedback loop oscillator, that was first developed by Stricker et al. \cite{Stricker2008}. Our model consists of six delay differential equations, that take into account the canonical components of gene expression: DNA promoter states, RNA expression and protein translation and stability. Using two delay terms to represent intermediate states, we can quantitatively reproduce the experimentally reported inducer-dependent oscillatory behaviour, based on a DNA copy number asymmetry of the positive and negative feedback loops. Based on this model, we redesign of the network to examine whether we can improve the tunability of the oscillator, based on modifications at the level of (a) DNA copy numbers, (b) RNA degradation rates and (c) protein production rates. 

\section{Model}
We have modeled the oscillator based on the coupled dynamics of (a) DNA transcription to RNA based on promoter state dynamics, (b) translation of RNA to protein and (c) protein folding (Fig. \ref{fig:schematic}). The derivation and assumptions of this reduced model are described in detail in Appendix \ref{A2}.

{\it Transcription} is modeled in terms of the dynamics of two mRNA species, encoding the AraC activator ($m_a$) and LacI repressor ($m_r$). The general equation of mRNA copies ($m_x$) for activator ($x=a$) and repressor ($x=r$) is: 
\begin{eqnarray}
        \Dot{m}_x = \frac{n_x b (1 + \alpha k_1  A_{\tau_1})}{(1 + k_1 A_{\tau_1})(1+ k_2 R_{\tau_2})^2} - k^x_{m} m_x \text{,}\\ \text{where } x \in \{a,r\} \nonumber 
    \label{eq:mr}
\end{eqnarray}
The two parts of the RHS of the equation represent the production and degradation terms. The feedback of activator dimers (A) and repressor tetramers (R) determine mRNA levels with the subscripts $\tau_1$ and $\tau_2$ indicating respective delays due to the formation of intermediate states. 
The terms $A_{\tau_1} \equiv A(t-\tau_1)$ and $R_{\tau_2} \equiv R(t-\tau_2)$, represent the equivalence of the number of protein dimers where $t$ is current time. 
The copy numbers of the genes encoding \textit{araC} and \textit{lacI}, $n_a$ and $n_r$ respectively, act as multiplication factors. We assume an asymmetry in plasmid copy numbers, i.e. $n_a \neq n_r$, based on the reported requirement of copy number differences of the genes in the original experiment by Stricker et al. \cite{Stricker2008}. Transcription rate is determined by $b$, the basal transcription rate when there is no activator bound to the promoter, $\alpha$ the multiplicative increase in transcription rate when activator (A) is bound and the degradation rates $k^a_m$ and $k^a_m$ of $m_a$ and $m_r$ respectively. 
The terms $k_1$ and $k_2$ are equilibrium binding rates of the A dimer and R tetramer binding to the promoter. These are modeled as variables based on a previously described expression \cite{Stricker2008}. The equilibrium binding rate of AraC binding to the promoter, $k_1$, depends on the concentrations of the inducers arabinose ([ara]) and IPTG ([iptg]) as:

\begin{equation}
	k_1 = k^{min}_1 + (k^{max}_1 - k^{min}_1)\frac{[ara]^{2}}{\mu^{2} + [ara]^{2}} .\frac{\nu^{2}}{\nu^{2} + [iptg]^{2}}
	\label{eq:k1}
\end{equation}

and the range of activator binding to the promoter is set by $k^{min}_1 = 0\text{ molecules}^{-1}$,   $k^{max}_1 = 1\text{ molecules}^{-1}$ and $\nu = 1.8$ mM is a scaling parameter. The LacI-promoter binding rate $k_2$ is determined by:

 \begin{equation}
 	k_2 = k^{min}_2  + (k^{max}_2 - k^{min}_2) \frac{\lambda^{2}}{\lambda^{2} + [iptg]^{2}}
	\label{eq:k2}
\end{equation}

where  while the repressor binding range is determined by $k^{min}_2 = 0.01\text{ molecules}^{-1}$,   $k^{max}_2 = 0.2\text{ molecules}^{-1}$ and $\mu = 2.5 \% $ (in \% w/v) and  $\lambda = 0.035$ mM is a scaling parameter. The cooperativity of inducer driven binding of the proteins to the promoter is modeled by a second order dependence of both $k_1$ and $k_2$ on [ara], [iptg], $\nu$ and $\lambda$.

The mRNAs are translated into unfolded proteins $A_{uf}$ and $R_{uf}$, the monomeric activator and repressor respectively. The dynamics of translation of the unfolded protein species ($X_{uf}$) is given by the general expression-

\begin{eqnarray} 
	\Dot{X}_{uf} = \sigma_X m_X - k_{f}  X_{uf} - k_X X_{uf}\text{,}\\ 
	\text{where } X_{uf} \in \{A_{uf}, R_{uf} \}  \nonumber 
	\label{eq:ufProt}
\end{eqnarray}

where $\rho_X$ is the translation rate of X that corresponds to $A_{uf}$ and $R_{uf}$. This is also referred to as the ribosome binding site (RBS) efficiency in the synthetic biology literature \cite{Levin-Karp:2013aa}. The folding rate  $k_f$ is constant and common to both proteins, while the degradation rate of the unfolded protein $k_X$ is variable due the ssrA tag that results in rapid degradation \cite{Stricker2008} by proteolysis mediated by ClpXp \cite{Cookson2011}. To account for the fact that the number of ClpXp molecules available to degrade proteins is limiting, the degradation rate for repressor is given as-
\begin{equation}
    k_R = \frac{\gamma_0}{c_0 + \Sigma P}
      \label{eq:Rprotdegr}
\end{equation}

while that for the activator is given as-
\begin{equation}
	k_A =g \cdot k_r
\end{equation}
Here, $\gamma_0$ is the maximal degradation rate, $c_0$ = 0.1 is the concentration of proteins at which the rate of ClpXP is half-maximal and $g$= 2.5 represents the differential degradation of two proteins, i.e. the activator is degraded faster than the repressor. The total copy number of all proteins in this system, $\Sigma P = A_{uf} + R_{uf} + A + R$, thus inversely determines the degradation rate.

\begin{figure}[b]
\includegraphics[width=0.48\textwidth]{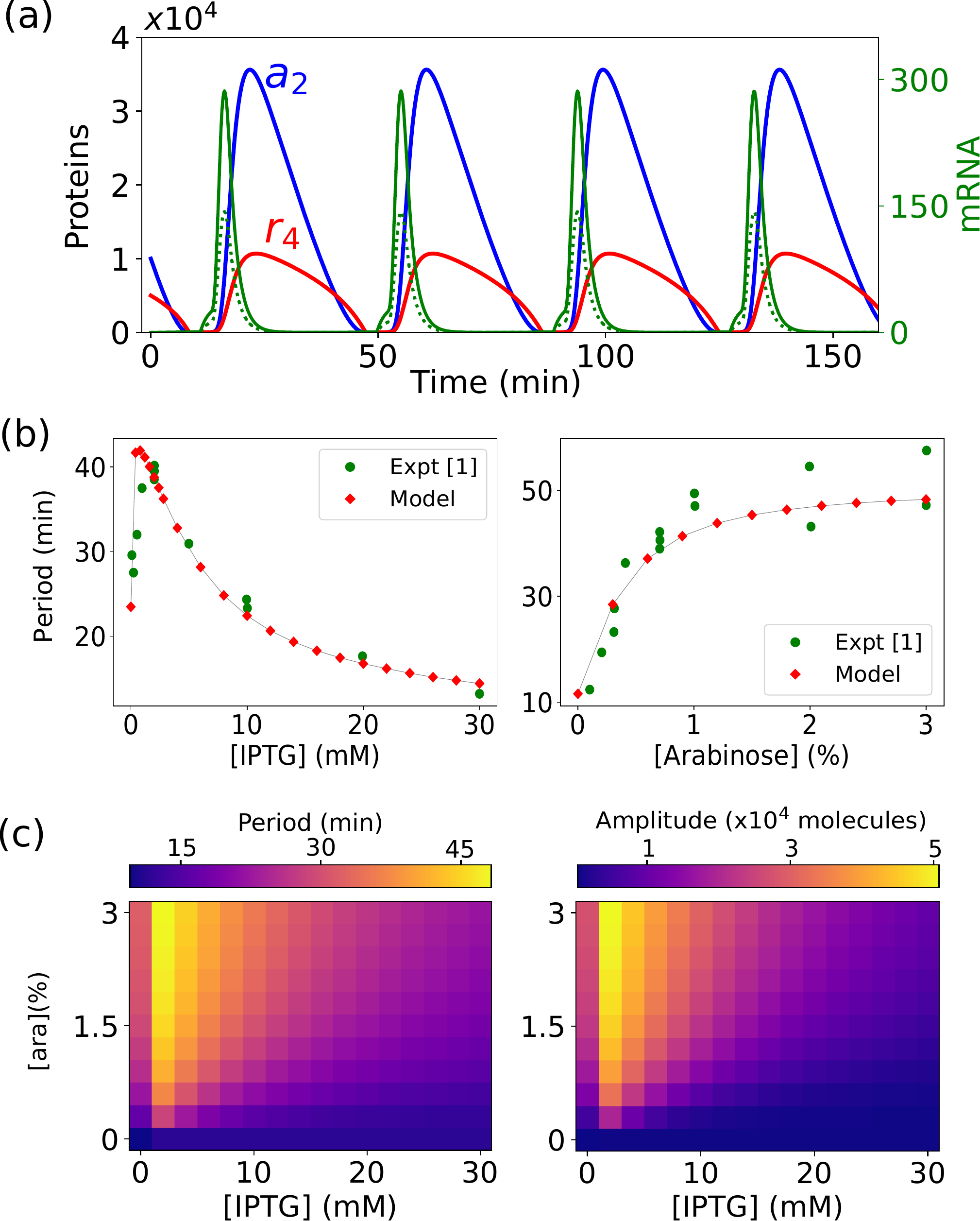}
\caption{(a) The number of molecules of AraC mRNA (green line) and protein dimers (blue line) and LacI mRNA (dotted green line) and protein tetramers (red line) 
are plotted as a function of time for the minimal model. Here, $n_a$=50, $n_r$=25, [ara]= 0.7\% and [IPTG]= 2 mM.  (b) The period of oscillations predicted by the model (red $\blacklozenge$) is compared to experimental data from Stricker et al. \cite{Stricker2008} (green $\bullet$) for increasing inducer concentrations of ({\it left}) IPTG ([arabinose]= 0.7\%) and ({\it right}) arabinose ([IPTG]=2 mM). (c) The phase diagram plots the effect of a systematic change in IPTG and arabinose concentrations on the oscillatory ({\it left}) period and ({\it right}) amplitude. Color bars represent the respective scales.}
\label{fig:Exptcomp}
\end{figure}

Dimers of AraC ($A$) and tetramers of LacI ($R$) are the active form of the proteins that regulate gene expression. We assume the monomers to be in rapid equilibrium with their respective dimers and tetramers, as a result of which the dynamics of AraC dimers are described by-
\begin{equation}
\Dot{A} = \frac{1}{p_a}k_{f} A_{uf} - k_a A 
\label{eq:aradimer}
\end{equation}
and the dynamics of LacI tetramers are described by-
\begin{equation}
	\Dot{R} = \frac{1}{p_r}k_{f} R_{uf} - k_r R
	\label{eq:lactetramer}
\end{equation}
The protein concentration is thus determined by folding rate $k_f$ of unfolded proteins ($A_{uf}$ and $R_{uf}$). 
The factors $p_a=2$ and $p_r=4$ account for the dimerization and tetramerization of the activator and repressor respectively, 
required to account for the rapid equilibrium of folded monomers with their multimers. 
As a consequence, at any time point, the dimers are half ($1/p_a$) the number of monomers and tetramers are 1/4th ($1/p_r$) the number of monomers at the same time. 


\begin{table}[h!]
    \centering
    \caption{The values of parameters used in the model are based on either previous reports \cite{Stricker2008}, or varied.}
    \begin{tabular}{p{1.75cm} p{3cm} p{6cm} p{2cm}} 
        Parameter & Value & Description & Reference \\ \hline 
        b & $0.36$ min$^{-1}$ & Basal transcription rate & \cite{Stricker2008} \\ 
        $\alpha$  & $20$ & Transcription activation & \cite{Stricker2008} \\
        $n_a$ & $50$ and varied & {\it araC} DNA copy number & \cite{Stricker2008}, this study  \\ 
        $n_r$ & $25$ and varied  & {\it lacI} DNA copy number & "  "\\ 
        $ k^a_{m}, k^r_{m}$  & $0.54$ min$^{-1}$ & Degradation rate of mRNA & "  "\\
        $\sigma_{A}$ & $90$ min$^{-1}$   & Translation rate (RBS efficiency) of {\it araC}& \cite{Stricker2008},  \\ 
        $\sigma_{R}$ & $90$ min$^{-1}$  & Translation rate (RBS efficiency) of {\it lacI}& \cite{Stricker2008}, this study \\ 
          $k_{f}$& $0.9$ min$^{-1}$  & Rate of folding of proteins & "   " \\ 
        $\gamma_0$  & $1080$ min$^{-1}$  & Maximal degradation & "  " \\
    \end{tabular}
    \label{tab:parm}
\end{table}

The simplifications of the model include assuming equilibrium of protein dimerization and tetramerization and promoter binding of the activator and inhibitor. Since each of these processes is expected to multiple intermediate states, we account for these in our equations by an explicit delay. Such an approach of using delay to account for intermediate processes (typically transcription) without explicitly modeling them has been used successfully used to model negative feedback oscillators \cite{Mather2009} and 
the lac operon \cite{Yildirim:2004aa}.

\section{Results}

\subsection{Minimal model can reproduce oscillatory dynamics from previous experiments} 

Our model is based on parameters taken from previous reports (Table \ref{tab:parm}), leaving only two free parameters, the delay terms $\tau_1$ and $\tau_2$ for the activator and repressor respectively. Since the activator $A$ must only transition through one step, dimerization, to be active, while the active repressor $R$ undergoes two intermediate reactions of dimerization and tetramerization \cite{Stricker2008}, we assume $\tau_2 = 2\tau_1$. 
Having reduced the free parameter to only one free parameter $\tau_1 $, we estimate the delay in monomer to dimer transition to be in the order of minutes, 
based on previous reports \cite{Mather2009}. We then numerically optimized $\tau_1$ by minimizing the deviation between predicted and experimentally reported period of the oscillations by Stricker et al. \cite{Stricker2008}, for [IPTG] = 2mM and 0.7\% arabinose. A scan of varying it from $\tau_1$ from 0 to 5 minutes with 0.25 step-size resulted in an optimal value of $\tau = 1.25$, at which the sustained oscillations in the concentrations of AraC and LacI mRNA and protein complexes are observed (Fig. \ref{fig:Exptcomp}(a)), with 0.7\% arabinose and 2 mM IPTG as inducer concentrations. The qualitative nature of the oscillations is comparable to that of the detailed model of Stricker et al. in previous work \cite{Stricker2008}. The oscillations show the same period and amplitude, independent of initial conditions. 

The so called `degrade-and-fire' (DF) behaviour of the oscillator observed \cite{Mather2009} is consistent with the mechanism of a basal mRNA transcription rate ($b$) producing low concentrations of protein, that are constantly degraded by proteolysis, determined by $\gamma_0$ and $c_0$. 
Since the production of {\it AraC} dimers is faster than the {\it LacI} tetramers, the activator protein induces network expression, initially at a slow rate, and then at increasing rates via the positive feedback loop (`fire'). The repressor protein concentration is however also gradually increasing, and once it reaches a threshold required for promoter repression, the negative feedback loop begins to inhibit transcription. This negative feedback reduces mRNA levels, and combined with the mRNA degradation rates $k_m^a$ and $k_m^r$, turns the network off (`degrade'). The basal transcription rate then once more induces the cycle, resulting in oscillations. 

We find our minimal delay differential equation model 
predicts that the time period of the oscillator responds to IPTG and arabinose induction differently (Fig. \ref{fig:Exptcomp} (b)). The period increases for 0 to 1 mM [IPTG] and then decrease for higher values when arabinose is constant (0.7\%). On the other hand, increasing arabinose concentrations from 0 to 3\%, with constant IPTG (2 mM) results in an increase and saturation of the period. Surprisingly, these model predictions closely match previously reported experimentally determined values \cite{Stricker2008}. Indeed both the period and amplitude of oscillations show similar behavior over wider ranges of both inducer concentrations (Fig. \ref{fig:Exptcomp} (c)).

Since the ability of the `fire, inhibit and degrade' model to produce oscillations depends on an asymmetry in activator and repressor molecules, we tested the sensitivity of the model to both the extent of asymmetry and whether DNA asymmetry can be mimicked by translational asymmetry.

%
%
%

\begin{figure}[t!]
\includegraphics[width=0.45\textwidth]{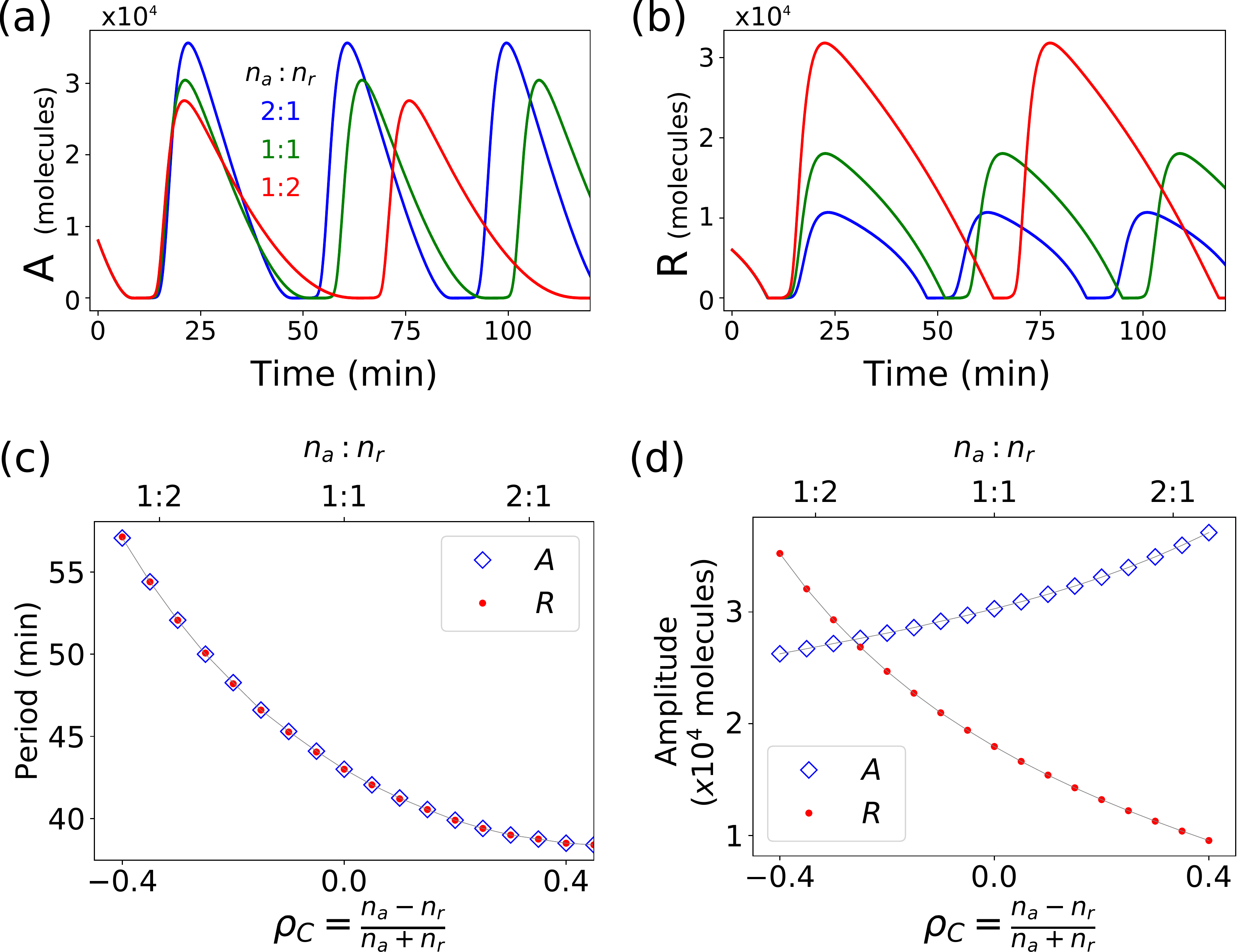}
\caption{{\bf Effect of relative DNA copy number asymmetry on oscillations.} (a, b) The effect of the gene copy numbers ratio of the activator $n_a$ ({\it araC}) and repressor $n_r$ ({\it lacI}) on the (a) activator dimer and (b) repressor tetramer concentration is plotted as a function of time. Individual plots signify $n_a:n_r$ 2:1 (blue), 1:1 (green) and 1:2 (red). 
(c,d) The effect of the copy number asymmetry ($\rho_C$) and ratio $n_a:n_r$ on (c) oscillatory period and (d) amplitude of the activator are plotted. [IPTG]= 2 mM, [Arabinose]= 0.7\%, $n_a$=50 and $n_r$= 19 to 150.} 
\label{fig:cpnvar}
\end{figure}

\subsection{Effect of gene copy number asymmetry on oscillations}
In previous experimental and theoretical work where the {\it lac-ara} oscillator was first developed \cite{Stricker2008}, the genes encoding the activator and repressor were expressed from two different plasmids that were maintained in cells at different copy-numbers. The gene copy numbers of DNA molecules encoding activator was 50 and repressor was 25, a ratio of $n_a:n_r=2:1$. Our model uses the same values in order to reproduce experimental data (Fig. \ref{fig:Exptcomp}). However, both cell to cell variability due to molecular states, as well the differences in the stage of division, could result in variations in plasmid copy numbers. Additionally, the precision of absolute plasmid copy number control cannot be ruled out. 
As a result, we have examined the effect of varying the asymmetry of the DNA copies of activator ($n_a$) and repressor ($n_r$) on the oscillatory dynamics. We find that when the activator gene copies are two-fold in excess of or equal to those of the repressor, the oscillations are rapid and amplitudes comparable. However, a two-fold excess of repressor copies results a small increase in the period of oscillation of both activator protein $A$ (Fig. \ref{fig:cpnvar}(a)) and repressor $R$ (Fig. \ref{fig:cpnvar}(b)). Indeed the asymmetry, $\rho_C =(n_a - n_r)/(n_a + n_r)$, appears to drive the increased amplitude of that gene, i.e. if the asymmetry involves an excess of {\it lacI} DNA, then the repressor protein $R$ has a higher amplitude than the activator $A$, and vice-a-versa.

A systematic scan across $\rho_C$ ranging from -0.4 to 0.4 (repressor DNA in excess to activator DNA in excess) demonstrates a continuous decrease in period, i.e. increase in oscillation frequency (Fig. \ref{fig:cpnvar}(c)), while repressor amplitude reduces and activator amplitude increases (Fig. \ref{fig:cpnvar}(d)). This would suggest that at some higher factor of asymmetry (e.g. $\rho_C=-1$), the oscillator period would become very large and not allow for `rapid' oscillations, within a generation of the bacterium. Indeed it suggests that a DNA copy number asymmetry ensures fast oscillations. If the oscillator were to be re-designed with an equal copy of both genes, we proceeded to ask if the asymmetry of mRNA translation could replace the asymmetry in DNA copies.

\begin{figure*}[ht!]
	\includegraphics[width=0.7\textwidth]{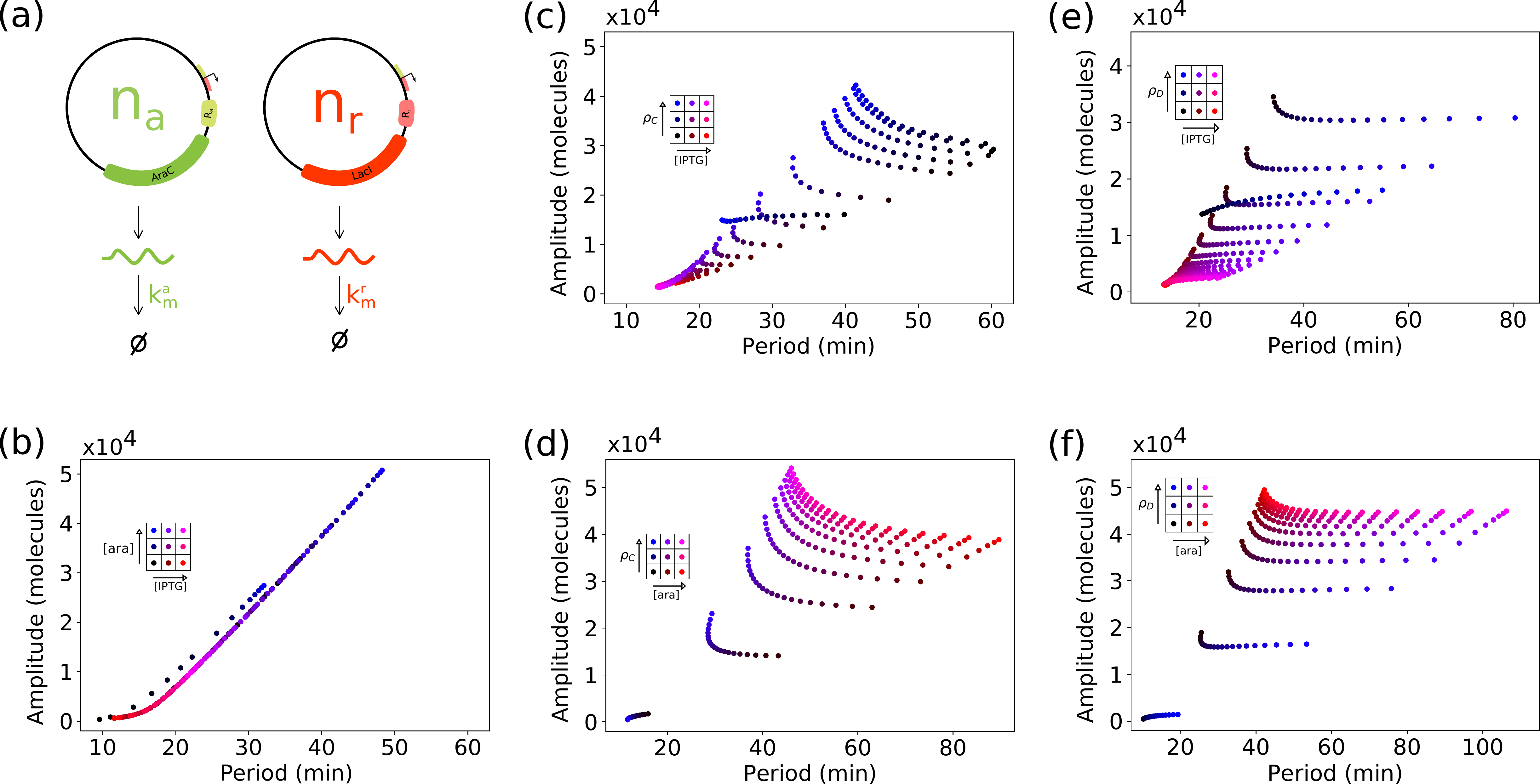}
\caption{{\bf Network tunability of oscillator with asymmetry in gene copies ($\rho_C$) and mRNA degradation ($\rho_D$).} (a) The schematic represents the asymmetry of activator and repressor that we test in terms of the copy numbers $n_a$ and $n_r$ ($\rho_C$), and the degradation rates $k_a$ and $k_r$ ($\rho_D$) respectively. (b-f) The tunability of oscillatory amplitude as a function of period based on (b) varying concentration of the inducers IPTG and arabinose when $\rho_C=0.33$ ($n_a =50$, $n_r=25$) is compared to (c,d) the effect of increasing copy number asymmetry in response to (c) either changing IPTG while [arabinose]= 0.7\% or (d) varying arabinose while [IPTG]= 2 mM. (e,f) A similar scan of parameters is performed for $\rho_D$ changes in response to increasing (e) [IPTG] or (f) [arabinose]. The colors in the plots indicate parameter values input: [IPTG]= 0 to 30 mM in steps of 2 mM, [arabinose]= 0 to 3\% in steps of 0.3\%, (3)$\rho_C$ and $\rho_D$= -0.5 to +0.5 in steps of 0.05.}
\label{fig:compTom}
\end{figure*}

\subsection{DNA copy number and RBS efficiency have an equivalent effect on protein oscillations}
We aimed to address the question whether a symmetric DNA copy number ratio ($\rho_C$) can be replaced with an equivalent mRNA translation asymmetry, based on ribosome binding site (RBS) affinity, where the RBS-based asymmetry is $\rho_R = (t_a - t_r)/(t_a + t_r)$. To answer this question, we attempt to derive an expression for protein translation in terms of the transcription and translation terms. We start from the transcription rate equation for the mRNA production of activator ($m_A$) and repressor ($m_R$) expressed in terms of the general equation for $m_X$ as-
\begin{align}
    \frac{dm_X}{dt} &=  n_Xg(A_{\tau_1}, R_{\tau_2})- k^X_m m_X, \quad X \in \{a,r\}
\end{align}
We now use the integrating factor method to write the functional form of the solution- 
\begin{align}
    m_X = e^{-k_mt} \int n_Xg(A_{\tau_1},R_{\tau_2}) e^{k_mt}dt 
\end{align}
where the notation of $k_m$ is assumed to correspond to $k_m^a$ and $k_m^r$. Substituting this in the protein translation equation that results in unfolded protein ($X_{uf}$) from mRNA, we obtain-
\begin{eqnarray}
    \frac{dX_{uf}}{dt} =&& \ \rho_X \Big( e^{-k_mt}\int n_Xg(A_{\tau_1},R_{\tau_2}) e^{k_mt}dt \Big) \nonumber \\
    &&- k_{f}  X_{uf} - k_X X_{uf} 
\end{eqnarray}

Since, $n_X$ is a constant, it is taken out of the integration resulting in the expression of the rate of unfolded protein formation-
\begin{eqnarray}
    \frac{dX_{uf}}{dt} =&& \ \big(\rho_X n_X \big)\Big( e^{-k_mt} \int g(A_{\tau_1}, R_{\tau_2}) e^{k_mt}dt \Big) \nonumber\\
    &&- k_{f}  X_{uf} - k_X X_{uf} 
\end{eqnarray}
This expression has a multiplicative factor consisting of two constants $n_X$ and $\rho_X$, demonstrating the exact equivalent of DNA copy number and RBS-efficiency. Since the production of RNA functions as a simple multiplicative factor of the DNA copy number, we further tested whether the degradation rate would potentially affect the dynamics any differently. 

\subsection{Robust design of oscillator based on mRNA degradation rate asymmetry} 
Based on these results, excessive expression of the activator appears to be important to ensure the rapid (small period) oscillations of the proteins with significant amplitude. However, since the DNA copy number asymmetry ($\rho_C$) based on plasmid copies can be subject to variability, an alternative means of generating this asymmetry needs to be considered. Since the effect of RBS efficiency on protein translation can be factorized in terms of DNA copy numbers (i.e. they have the same effect on oscillations), we proceeded to explore the role of mRNA degradation rate asymmetry $\rho_D = (k^a_m - k^r_m)/(k^a_m + k^r_m)$, as seen in Fig. \ref{fig:compTom}(a), on oscillations. 

We find the effect of increasing IPTG concentration result in an almost linear decrease in period of oscillations (higher frequency), while increasing arabinose concentrations increases the amplitude, for a fixed DNA copy numbers asymmetry of $\rho_C$=0.33, i.e. $n_a$ is 50 and $n_r$ is 25 (Fig. \ref{fig:compTom}(b)). If on the other hand, the $\rho_C$ is itself varied, then we find a dependence of oscillation period varies on the asymmetry, for a given IPTG or arabinose concentration (Fig. \ref{fig:compTom}(c,d)). If we now vary the mRNA degradation rate asymmetry $\rho_D$, we find the oscillator explores a wider period (20-100 min) as compared to the copy-number asymmetry (Fig. \ref{fig:compTom}(e),(f)). This would appear to suggest that while the experimental system based on two plasmids can produce fast, tunable oscillations in proteins, their robustness to cell to cell variability can be improved by expressing the proteins at the same level with an asymmetry in activator and repressor introduced in terms of mRNA degradation rates, that result in lifetime differences.

\section{Discussion}
We have successfully developed a minimal model of the dual feedback loop tunable genetic oscillator, which still maintains sufficient detail to explore the regulation of the system at the levels of the canonical central dogma- DNA, RNA and protein. We have optimized a single free parameter, the delay term to obtain oscillatory dynamics that match the whole range of experimental values. Using this minimal model, we explore the role of asymmetry of activator and repressor, in terms of DNA copy number and the production and degradation rates of mRNA. We find that while rapid oscillations can be produced when the a two-fold excess of activator genes is present, this asymmetry is mathematically identical to mRNA production rates. On the other hand, mRNA degradation rate asymmetry results in a wider tunability of the oscillator. Thus, we believe with this simplified model of a canonical synthetic genetic oscillator, we predict a more robust design of the oscillator, capable of exploring a wider range of frequency and amplitudes, than previously described.

While attempts to simplifying the lac-ara dual-feedback oscillator have been made, they have ignored the role of mRNA modeling only the DNA and protein components \cite{Veliz-Cuba2015,Tomazou2018}. As a result, the ability to explore novel network design is limited. In their work, multiple parameters are fit to experimental data, which could result in artifacts. In our work we have maintained identical parameters to the published model and experimental work of Stricker et al.  \cite{Stricker2008}, as well as used experimentally tested ranges of inducers (IPTG and arabinose), while leaving only one free parameter, the delay term. We believe thus to have improved upon previous attempts at simplification of this model oscillatory network. 

While we have explored the role of asymmetry in oscillatory dynamics, it is clear from our simulations that copy number and translation symmetry can also result in oscillations. This is consistent with the results of a model of where dual-feedback oscillator with symmetric components can still oscillate as described by Maeda et al. \cite{Maeda2012}. Indeed a comparison of genetic oscillator designs suggests that symmetric gene networks while used in models are unlikely to be natural \cite{Purcell2010}. Indeed the assumption of many standard genetic oscillator models such as the `repressilator' model is that mRNA degradation rates are comparable for the component genes \cite{Elowitz2000}. Indeed recent work has demonstrated the possibility to achieve fine-tuned control over protein levels through mRNA degradation rate modulation, simply by varying the length of the polyA tail \cite{Arthur:2017aa}. Here by exploring the possible differences in copy numbers, RBS efficiency and mRNA stability of the two-component network, we have explored more biologically realistic properties of networks, and find the effects of asymmetry depend on the level in the information flow at which they are implemented- DNA, RNA or protein. 

The role of the host cell machinery in determining aspects of the genetic oscillator dynamics has been explored previously showing negative-feedback loop oscillators such as circadian oscillators can phase lock to the cell cycle \cite{Paijmans:2016aa}, 
which natural oscillators avoid by using protein-phosphorylation, asynchronous replication and `noise \cite{Paijmans:2017ab}. This would suggest the robustness of the lac-ara oscillator seen in experiments could be the combined result of the positive feedback loop as well as the noise coming from replication asynchrony and copy number variability \cite{Stricker2008}. Experiments with the oscillator genes are localized on the same plasmid, to separate the effects due to copy number variability and other sources of noise, could potentially answer some of these questions.

\section*{Acknowledgments}
We wish to acknowledge discussions with Dr. Pranay Goel regarding delay differential equations. YKJ is supported by a project assistantship from a grant BT/PR16591/BID/7/673/2016 from the Dept. of Biotechnology, Govt. of India to CAA.

\section{Appendix}

\subsection{Numerical Simulations \& Analysis}
In summary, we have reduced the mathematical model given by Ref. \cite{Stricker2008}, with more than 30 equations, to a simplified model having only six delay differential equations. We use two delay terms, $\tau_1$ (activator) and  $\tau_2$ (repressor) and relate them as $\tau_2 = 2\tau_1$. 
This is based on the fact that activator proteins bind to DNA after dimerization, a 1-step process, but the repressor binds after tetramerization, involving two intermediate steps. Since the rate constants for all these reactions have the same value according to previous reports \cite{Stricker2008}, and  promoter binding is much faster than multimerization, 
we argue that the delay due to two consecutive reactions is twice that of a single reaction. 
This model was simulated using Python 3.6 and for the purpose of solving delay differential equations, PyDDE package was used. To solve a delay differential equation numerically, the solver needs to be given initial data of all the variables between time $0<t<2\tau$. To get around this problem, we solve the equations as simple ODEs between $0<t<2\tau$ and as DDEs when $t> 2\tau$ 

A custom written peak-finding algorithm was used for estimation of time period and amplitude. This algorithm involved comparing the concentration value at each time point with that of some time before and some time after the particular point. If $y(t)$ denotes the time series of protein concentrations, for any time $t$, we check if-
\begin{equation}
    y(t_0) - y(t_0-s) > 0 \ \ \ \text{and} \ \ \  y(t_0) - y(t_0+s) > 0
\end{equation}
If the above condition is satisfied for all $s \ \in  \ \ \{ \delta t, 2\delta t, 30\delta t \}$
then we say that $t_0$ is a `peak'. Our calculations were run with $\delta t$ values of 0.2 min. The period and amplitude can be found by averaging the time difference between consecutive peaks and the values of $y(t_0)$, respectively.

\subsection{Model derivation}\label{A2}
We have derived the model equations by simplifying the detailed reaction kinetics of (a) promoter dynamics determined by protein binding, (b) RNA transcription, (c) translation of mRNA and (d) protein folding and (e) protein multimerization. The promoter reactions in the AraC-LacI oscillatory system are as follows- 
\[P^{a/r}_{0,j} \ + \ A \xrightleftharpoons{k_1} \ P^{a/r}_{1,j} \]
\[P^{a/r}_{i,0} \ + \ R \xrightleftharpoons{2k_2} \ P^{a/r}_{i,1} \]
\[P^{a/r}_{i,1} \ + \ R \xrightleftharpoons{\frac{1}{2}k_2} \ P^{a/r}_{i,2} \]
where $A$ represents AraC protein dimers, $R$ represents LacI protein tetramers, $P^{a/r}_{i,j}$ represent the states of promoters on the (a)ctivator/(r)epressor plasmids with $ i \in (0,1) $ number of AraC dimers bound and $j \in (0, 1, 2)$ the number of LacI tetramers bound. In contrast to the previous work \cite{Stricker2008}, we consider$k_1$ and $k_2$ to be the equilibrium constants, and equal to the ratio of rate of forward reaction to the rate of backward reaction. 
The mRNA transcription reactions are- 
\[ P^{a/r}_{0,0}  \xrightarrow{b_a} \ P^{a/r}_{0,0} + m_{a/r}\]
\[ P^{a/r}_{1,0}  \xrightarrow{\alpha b_a} \ P^{a/r}_{0,1} + m_{a/r}\]
 where $m_{a/r}$ represents the number of mRNA molecules of {\it araC}/{\it lacI} genes. When an AraC dimer is bound to the promoter, i.e. when promoter is in state $P^{a/r}_{1,0}$, transcription rate is increased by a factor of  $\alpha$. When any number of LacI tetramers are bound, i.e $ P^{a/r}_{i,1}$ or $P^{a/r}_{i,2}$, there is no transcription.  

The reactions representing mRNA translation and protein folding are-
\[ m_a \xrightarrow{t_a} m_a + A_{uf}\]
\[ m_r  \xrightarrow{t_r} m_r + R_{uf}\]
\[A_{uf}  \xrightarrow{k_{fa}} A\]
\[R_{uf}  \xrightarrow{k_{fa}} R\]
$A_{uf}$ and $R_{uf}$ are the unfolded activator and repressor proteins, $A$ is the dimeric AraC protein and $R$ is the tetrameric repressor protein. Here, unlike in previous work, we have assumed rapid equilibrium between folded monomer proteins, dimers and (in the case of LacI) tetramers.

Degradation reactions for molecular components- 
\[m_a \xrightarrow{\gamma_{ma}} \phi \]
\[m_r \xrightarrow{\gamma_{mr}} \phi \]
\[A_{uf} \xrightarrow{k_a} \phi \]
\[R_{uf} \xrightarrow{k_r} \phi \]
\[A \xrightarrow{k_a} \phi \]
\[R \xrightarrow{k_r} \phi \]
We have ignored protein looping, which is considered by Stricker et al. (2008) in their model.

Based on previous work where equilibrium assumptions were made while modeling genetic networks  \cite{Hasty2001,Hasty2002}. Every plasmid has one promoter and that promoter can be in total six possible states. The different promoter states bound to the transcription factor dimers and tetramers is given in the following equilibrium equations-

\begin{align}
\begin{split}
P_{1,0} & = k_1  A_2 P_{0,0} \\  
P_{0,1} & = 2 k_2 R_4 P_{0,0}  \\ 
P_{0,2} & = \frac{k_2}{2} R_4 P_{0,1} = \frac{k_2}{2} R_4  \Big( 2 k_2 R_4 P_{0,0} \Big) = k^2_2 R^2_4 P_{0,0} \\ 
P_{1,1} & = \Big( k_1 A_2 \Big) P_{0,1} = 2 k_1 A_2 k_2 R_4 P_{0,0} \\ 
P_{1,2} & = \Big( k_1 A_2 \Big) P_{0,2} = k_1 A_2 k^2_2 R^2_4 P_{0,0} 
\end{split}
\label{s2}
\end{align}
Here $A_2$ refers to AraC dimers (we refer to as $A$) and $R_4$ refers to LacI tetramers (we refer to as $R$). We have suppressed the promoter superscript $a$ and $r$ to simplify notation, but the following equations are independently valid for the promoter upstream of activator and repressor. Since the total number of all promoter states added together is constant and determined by the plasmid copy number $n$, so we impose the condition-  

\begin{equation}
\begin{split}
n = \sum_{i,j}P_{i,j} &= P_{0,0} + 2 k_2 r_4 P_{0,0} +  \Big( k_2 r_4 \Big)^2 P_{0,0} +  k_1 a_2 P_{0,0} + \Big( k_1 a_2 \Big) \Big( 2 k_2 r_4 \Big) P_{0,0} + \Big( k_1 a_2 \Big)\Big(k_2 r_4\Big)^2 P_{0,0}  \\
&= \bigg[ 1 +  k_1 a_2 \bigg] \bigg[1 + 2 k_2 r_4 +  \Big( k_2 r_4 \Big)^2 \bigg] P_{0,0} = \bigg[ 1 +  k_1 a_2 \bigg] \bigg[1 + k_2 r_4 \bigg]^2 P_{0,0}
\end{split}
\label{s3}
\end{equation}

The production of mRNA is governed by the differential equations - 

\begin{eqnarray}
\frac{dm_a}{dt} = b(P^{a}_{0,0} + \alpha P^{a}_{1,0}) -  \gamma_{m} m_a \\
\frac{dm_r}{dt} = b(P^{r}_{0,0} + \alpha P^{r}_{1,0}) -  \gamma_{m} m_r
\label{Seq:rna}
\end{eqnarray}
Rewriting $P_{1,0}$ in terms of $P_{0,0}$ using \ref{s2}, and then writing $P_{0,0}$ in terms of $n$ using Eq. \ref{s3}, into the above expression (Eq. \ref{Seq:rna}) we get the following differential equation- 
\begin{align}
\begin{split}
\frac{dm_a}{dt} = \frac{b(1 + \alpha k_1 a_2 )}{\big( 1 +  k_1 a_2 \big) \big(1 + k_2 r_4 \big)^2} -  \gamma_{m} m_a \\
\frac{dm_r}{dt} = \frac{b(1 + \alpha k_1 a_2 )}{\big( 1 +  k_1 a_2 \big) \big(1 + k_2 r_4 \big)^2} -  \gamma_{m} m_r
\end{split}
\end{align}
Even though we have almost derived the equation stated in the main equation, it remains to determine $A_2$ and $R_4$, which can only come from the protein production equation. Hence, the feedback present in the system is also mimicked in our model. To proceed towards the dynamics of protein dimers and tetramers required, we must first describe the intermediate dynamics of newly translated but as yet unfolded proteins-
\begin{align}
\begin{split}
\frac{dA_{uf}}{dt} &= \rho_a m_a - k_{f}  A_{uf} - k_a A_{uf} \\
\frac{dR_{uf}}{dt} &= \rho_r m_r - k_{f}  R_{uf} - k_r R_{uf}     
\end{split}
\end{align}
Here, the first term represents the production of unfolded proteins by translation,  the second and third term represent depletion of proteins due to folding and degradation respectively. We can now write a reaction for protein monomer production as-
\begin{align}
\begin{split}
\frac{dA'}{dt} &= k_{f} a_{uf} - k_a a \\
\frac{dR'}{dt} &= k_{f} r_{uf} - k_r r 
\end{split}
\end{align}
where $A'$ and $R'$ are the numbers of monomeric activator and repressor proteins produced by translation, respectively. Our aim is to describe the dynamics of protein dimers and tetramers. We again use the equilibrium assumption for this purpose. Let the number of monomers of AraC and LacI proteins remaining in the system after dimerisation and tetramerisation be $A'$ and $R'$ for activator and repressor respectively and the number of dimers of AraC and tetramers of LacI thus formed be $A_2$ and $R_4$ respectively. 
Since the total number of protein molecules ($A_{tot}$, $R_{tot}$) is conserved, we have-
\begin{align}
A_{tot} &=  A' + 2  A_2  \approx 2  A_2 \ \ \ \Rightarrow  \ \ \  A_2 = \frac{A'}{2}\\
R_{tot} &=  R' + 2  R_2 + 4  R_4 \approx  4  R_4 \ \ \ \Rightarrow \ \ \   R_4 = \frac{R'}{4}
\end{align}
where all the components are in `monomer equivalents'. A differential equation for active forms of the proteins can be written for activator dimers as-

\begin{align}
\begin{split}
	\frac{d A_2}{dt} = \frac{d A_2}{dA} \frac{dA}{dt} = \frac{1}{2} \frac{dA}{dt} &= \frac{1}{2} k_{f} A_{uf} - \frac{1}{2} k_a A'  \\ 
&= \frac{1}{2} k_{f} A_{uf} - k_a  A_2 
\end{split}
\end{align}
and for repressor tetramers as-
\begin{align}
\begin{split}
\frac{d R_4}{dt} = \frac{d R_4}{dr} \frac{dR'}{dt} = \frac{1}{4} \frac{dR'}{dt} 
	&	= \frac{1}{4} k_{f} R_{uf} - \frac{1}{4} k_r R'  \\ 
	&	= \frac{1}{4} k_{f} R_{uf} - k_r  R_4
\end{split}
\end{align}

To simplify the notation, we again redefine $A =  A_2$ and $R=  R_4$. We also introduce delay into the first equation heuristically, using the argument that the intermediate equilibrium reactions of dimerisation and tetramerization take some time. The kinetics are given by the forward rates of dimerization of activator ($k_{d}^{a}$) and repressor  ($k_{d}^{r}$) and tetramerisation of repressor ($k_{t}^{r}$)  The promoter binding reactions may also contribute to the delay. These rates are taken from the previous study by Stricker et al. \cite{Stricker2008} are- 

\begin{table}[h!]
    \centering
    \begin{tabular}{p{2cm} p{2cm} p{5cm}}
    \hline
       {\bf Parameter} & {\bf Value} & {\bf Description} \\ \hline
        $k^b_1$  & 1.8 $s^{-1}$ & Backward rate of promoter binding binding by AraC/LacI \\
        $k_{d}^{a},k_{d}^{r},k_{t}^{r}$  & 0.018 molecules$^{-1}\cdot s^{-1}$ & Forward rates of dimerisation and tetramerisation \\
        $k_{-d}^{a},k_{-d}^{r},k_{-t}$ & 0.00018 $s^{-1}$ & Backward rates of dimerisation and tetramerisation\\ 
    \end{tabular}
    \caption{Rates of some equilibrium reactions are taken from Stricker et al. \cite{Stricker2008}}
    \label{tab:parmsup}
\end{table}
Thus, given the slow rate of dimerisation and tetramerisation compared to promoter binding rates (Table \ref{tab:parmsup}), we are justified in the assumption that all the delays in our equations are due to the protein multimerization. Based on the identity of values of $k_{d}^{x}$ and $k_{t}^{r}$, the delay in the feedback of the tetramer is twice that of the delay in the dimer. This allows us to arrive at six equations that fully describe the system, as described in the main text.

\bibliography{OscillatorLit}
\bibliographystyle{plain}
---
\beginsupplement
\clearpage

\section{Supplementary material}
\begin{figure}[h!]
\includegraphics[width=0.7\textwidth]{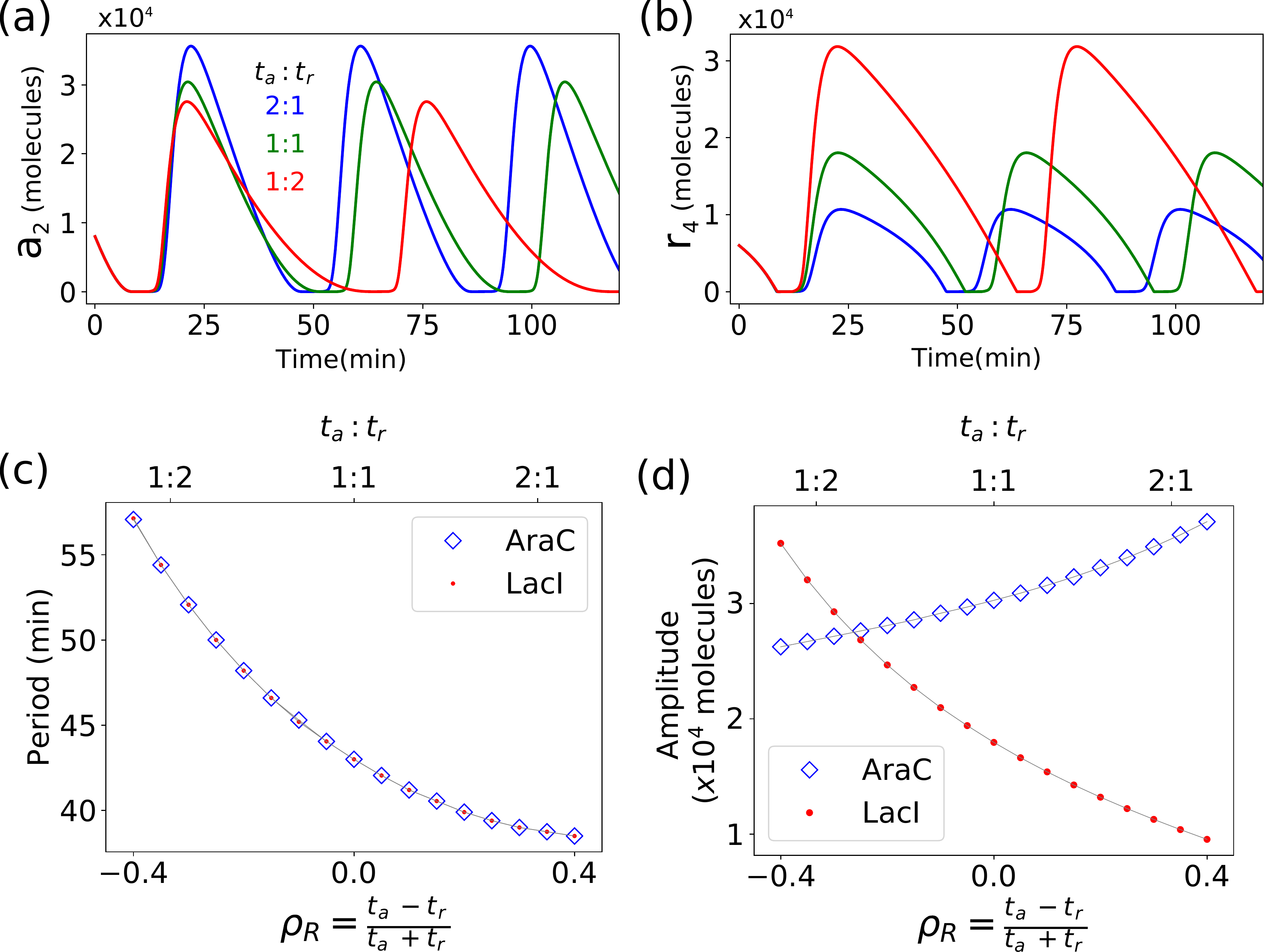}
\caption{Oscillations in the concentration of (a) AraC dimers and (b) LacI tetramers are obtained even when the copy number ratio ($n_a:n_r$) is kept fixed at (1:1) and RBS efficiency ratio ($t_a:t_r$) is varied, simulations were run for constant IPTG and arabinose input of 2 mM and 0.7\% respectively. For the same IPTG and Arabinose concentration, the dependence of (c) time period and (d) amplitude on the copy number ratio is shown. It can be seen that this produces exactly the same effect as keeping the RBS efficiency constant and changing the copy number}
\label{fig:3mod}
\end{figure}

\begin{figure}[h!]
\includegraphics[width=0.48\textwidth]{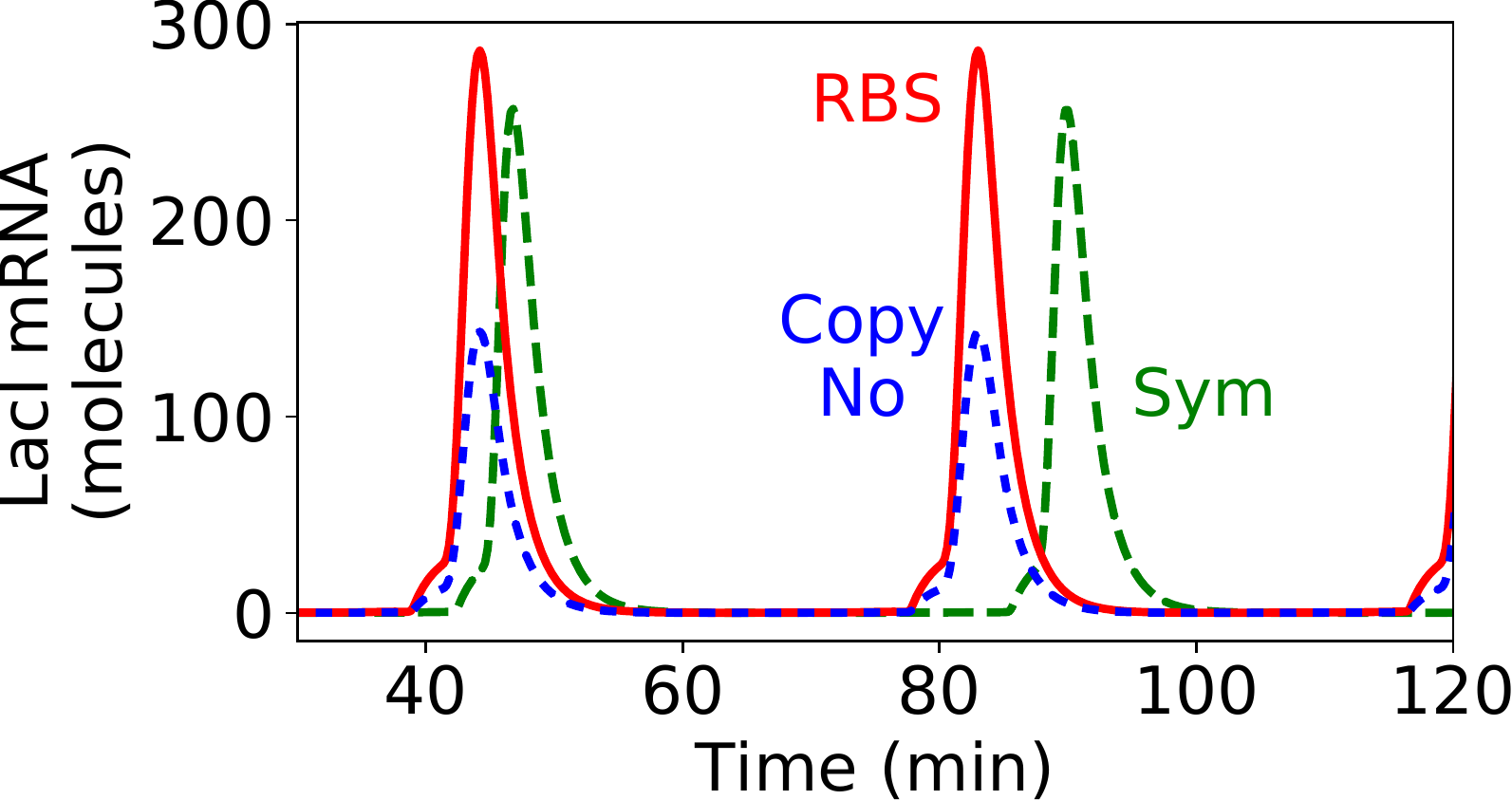}
\caption{number of LacI mRNA molecules present in the system for three different sets of parameter values- Copy no asymmetry($n_a/n_r = 2, t_a/t_r =1$), RBS asymmetry($n_a/n_r = 1, t_a/t_r =2$) and Symmetry ($n_a/n_r = 1, t_a/t_r =1$). This representative figure shows that even though protein dynamics are the same for both Copy number asymmetry model and RBS asymmetry model, mRNA dynamics are different and the two systems are not trivially equivalent. }
\label{fig:3mod}
\end{figure}
\end{document}